\documentclass[aps,twocolumn,showpacs,amsmath,amssymb]{revtex4}

 \usepackage{xcolor}

\usepackage{graphicx}
 \usepackage[colorlinks=true, linkcolor=green, citecolor=blue]{hyperref}

 \def\T{\textstyle}
 \def\l{\left}
 \def\r{\right}
 \def\nf{n_{\!f}}
 
 \def\ie{i.\,e.}

 \def\be{\begin{equation}}
 \def\ee{\end{equation}}
 \def\bea{\begin{eqnarray}}
 \def\eea{\end{eqnarray}}
 \def\bean{\begin{eqnarray*}}
 \def\eean{\end{eqnarray*}}
 \def\gsim{\mathrel{\rlap{\lower0.2em\hbox{$\sim$}}\raise0.2em\hbox{$>$}}}
 \def\ksim{\mathrel{\rlap{\lower0.2em\hbox{$\sim$}}\raise0.2em\hbox{$<$}}}
  \def\kg{\mathrel{\rlap{\lower0.25em\hbox{$>$}}\raise0.25em\hbox{$<$}}}

 \def\bm#1{\mbox{\boldmath$#1$}}
 \newcommand{\eq}[1]{(\ref{#1})}
\begin{document}

\title{Turning on the Charm}

\author{Andr\'e Peshier}
\affiliation{SUBATECH, Universit\'e de Nantes, EMN, IN2P3/CNRS
\\ 4 rue Alfred Kastler, 44307 Nantes cedex 3, France}

\date{\today}

\begin{abstract} \noindent
We argue that the strong jet quenching of heavy flavors observed in heavy-ion collisions is to a large extent due to binary scatterings in the quark-gluon plasma. It can be understood from first principles: the charm collision probability beyond logarithmic accuracy and Markov evolution.
\end{abstract}


\maketitle

In particle physics, heavy flavors are suitable to investigate properties of the strong interaction. Their large mass also proves useful when studying aspects of many-body QCD for heavy-ion phenomenology, since quarks with $m \gg T$ can be considered as {\sl test particles} in a heat bath.
Given this simplification, heavy quarks (say charm with $m_c \approx 1.2\,$GeV, which might be sufficiently large for temperatures reached at RHIC) can help to clarify the ongoing debate on the  parton energy loss in the quark-gluon plasma.
For {\sl light} partons, it is common to attribute jet quenching entirely to radiative energy loss, motivated by its parametric energy dependence \cite{Baier:1996sk, Gyulassy:2000fs}. One may note that plasma parameters inferred in such approaches can be hard to reconcile with general expectations, cf.\ \cite{Baier:2001yt}.
What is more,  the {\sl purely}  radiative picture of jet suppression has recently been challenged by the observation that heavy quarks (which radiate less \cite{Dokshitzer:2001zm}) are quenched almost as much as light quarks, as concluded by analyzing electron yields from heavy flavor decays \cite{Abelev:2006db, Adare:2006nq}.
This has revived an interest \cite{Mustafa:2003vh, vanHees:2004gq, Moore:2004tg, Wicks:2005gt} for the collisional energy loss as an additional suppression mechanism in the range of moderately large momenta. 

With regard to the importance of the understanding of jet quenching, it seems necessary to point out that existing approaches suffer from several fundamental shortcomings, which makes their conclusions evasive.
A majority of approaches is based on Fokker-Planck equations, despite the fact that the assumed dominance of soft scattering \cite{LL} is justified only at leading logarithmic accuracy \cite{Peigne:2007sd}. As one consequence, detailed balance must be {\sl imposed} in some way, whereas equilibration should be a prediction of the formalism.
A second severe issue of existing approaches is the disregard of the momentum dependence of the strong interaction. Strictly speaking, QCD calculations are not predictive without discussing the running coupling 
(opposed to QED with $\alpha_{_{\rm QED}} \approx 1/137$ in the widely applicable Thomson limit).
For quantitative estimates, a value of the coupling has then to be {\sl assumed}; common is $\alpha_{\rm fix} = \alpha(-(2\pi T)^2)$. 
However, the energy loss probes also parametrically soft scales, $\sim \sqrt\alpha\, T$, which is not only decisive for its high-energy limit \footnote{For $E \to \infty$, the energy loss from Eq.~\eq{eq:dEdx} is of {\sl first} order in the coupling \cite{Peshier:2006hi}, notwithstanding $|{\cal M}_i|^2 \sim \alpha^2$.}; it will also turn out to be crucial for heavy ion phenomenology.

\smallskip

Let us start by considering the average energy loss of a charm quark subject to binary collisions with gluons and light quarks ($i = g,q$) in the thermalized plasma,
\bea
	\frac{dE_i}{dx}
	&=&
	\frac{(dv)^{-1}}{2E}
	\int_k \frac{n_i(k)}{2k}
	\int_{k'} \frac{\bar n_i(k')}{2k'}\,
	\int_{p'} \frac1{2E'}
	\nonumber \\
	&& \hskip -7mm
	\times\, (2\pi)^4\delta^{(4)}(P\!+\!K\!-\!P'\!-\!K') \sum \l|{\cal M}_i\r|^2\, \omega \, .
 \label{eq:dEdx}
\eea
Here $d = 6$ for a charm quark; $v = p/E$ is its velocity, $n_i(k) = 1/(e^{k/T} \mp 1)$ and $\bar n_i = 1\pm n_i$ are the distribution functions of the collision partners, and $\omega = E-E'$ is the energy transfer in the scattering.
$\l|{\cal M}_i\r|^2$ is summed over color and spin states of all particles. 

Evaluating Eq.~\eq{eq:dEdx} with the Born cross sections \cite{Combridge:1978kx} would yield divergent results because of long-range gauge interactions. Therefore, screening effects (which formally arise from thermal loop corrections) have to be taken into account already in a leading order calculation. 
These thermal corrections come along with the vacuum fluctuations, which diverge and need to be renormalized. 
Renormalizability, as such, is obviously not affected by the (UV-finite) thermal contributions -- which might be a reason why renormalization is often utterly disregarded in finite-temperature field theory.
However, {\sl only} by this fundamental concept QCD calculations can be predictive.

Rigorous renormalization can be tedious, especially in thermal field theory. 
Fortunately, for the observables of interest here, the leading-order results can be inferred by elementary reasoning.
For gluon exchange processes ($t$-channel scattering) it has been argued previously \cite{Peshier:2006hi} that resumming and renormalizing the loop corrections amounts to replacing, in the Born amplitudes, the bare coupling by the running coupling $\alpha(t)$, schematically
\be
	\frac\alpha t \to \frac{\alpha(t)}{t-\Pi_T(\omega, q)} \, ,
\ee
where $q = (\omega^2-t)^{1/2}$.
We will parameterize the thermal self-energy $\Pi_T$ by an {\sl effective} cut-off, of the order of the Debye mass and evaluated with running coupling \cite{Peshier:2006ah},
\be
	\mu^2(t)
	=
	\kappa \cdot 4\pi \l( 1+\T\frac16\, \nf \r) \alpha(t) \, T^2 \, .
	\label{eq: mu2}
\ee
The customary {\em ad hoc} {\sl choice} for this screening mass is $\mu_{\rm fix}^2 = 4\pi \l( 1+\T\frac16\, \nf \r) \alpha_{\rm fix}\, T^2$, which we adopt only for the sake of comparison with existing estimates.
In fact, the coefficient $\kappa$ in Eq.~\eq{eq: mu2} can be calculated by comparing $dE^{t-{\rm channel}}/dx$, evaluated with effective cut-off, to the strict result beyond logarithmic accuracy \cite{PP}, which yields
\be
	\kappa = (2e)^{-1} \approx 0.2 \, .
	\label{eq: kappa}
\ee
The resulting cut-off is small, which will be a main source of differences between our and prevalent estimates.

Complementing the arguments put forward in \cite{Peshier:2006hi}, we now turn to the $s$ and $u$-channel contributions to charm scattering.
In order to quantify `the' coupling, consider the vacuum corrections to the Born amplitudes, which have logarithms of characteristic momenta. 
By specifying a renormalization scale $\mu_R$, bare quantities are expressed in terms of physical ones, such as the coupling $\alpha(\mu_R^2)$. 
By renormalization flow equations, all choices of $\mu_R$ are physically equivalent. 
This allows, if there is only {\sl one} characteristic momentum $P$, to choose $\mu_R=P$. Then the logarithmic term vanishes, and the renormalized result looks like the Born approximation -- albeit with $\alpha \to \alpha(P^2)$.
Now, the collisional energy loss \eq{eq:dEdx}, as a thermal average of $|{\cal M}_i|^2$, is dominated by interactions with $s \sim ET \gg |V^2|$, where $V^2 = t$ or $u-m^2$ \cite{Peigne:2007sd}.
For this particular kinematics, there is indeed only one relevant momentum scale determining $\alpha$ -- the virtuality $V^2$ of the intermediate state, cf.\ \cite{PeskinSchroeder}. We will fix the coupling also for the remaining {\sl sub-dominant} scattering contributions at the respective virtuality, not without estimating the arising uncertainty.
For time-like contributions, we continue the 1-loop coupling according to Ref.~\cite{Dokshitzer:1995qm},
\be
   \alpha(Q^2)
   =
  \frac{4\pi}{\beta_0}
   \l\{ \begin{array}{lc}
   L_-^{-1}
   \\[-0.35em]
   &   \ {\rm for \ } Q^2 \kg 0 \, ,
   \\[-0.35em]
   \frac12 - \pi^{-1} {\rm atn}( L_+/\pi )
   \end{array} \r.
   \label{eq: alpha(Q2)}
\ee
where $\beta_0 = 11-\frac23\, \nf$ with $\nf=3$, and $L_\pm = \ln(\pm Q^2/\Lambda^2)$.
With regard to quantitative estimates it is worthwhile recalling that perturbative approaches can be of use at surprisingly soft momentum scales \cite{Dokshitzer:2003qj}.
May details of the behavior of $\alpha(Q^2)$ in the deep infrared be uncertain, there exists a robust constraint (universality hypothesis),
\be
	\bar\alpha
	=
	Q_u^{-1}\int_{|Q^2| \le Q_u^2} dQ\, \alpha(Q^2)
	\,\simeq\,
	0.5 \, ,
\ee
where $Q_u = 2\,$GeV \cite{Dokshitzer:2003qj}. Accordingly, we impose an upper bound on the running coupling \eq{eq: alpha(Q2)}, $\alpha(Q^2) \le 1.1$ for our preferred QCD parameter $\Lambda = 0.2\,$GeV (adjusted to lattice results for the heavy quark potential, cf.~\cite{Peshier:2006ah}).
We have verified that the deep infrared region as well as the precise value of $\Lambda$ are not very important for our concerns -- for $s$ and $u$-channel terms due to phase space suppression, and for $t$-channel contributions by screening.

\begin{figure}[t]
	\vskip -2mm
	\centerline{\includegraphics[width=6.3cm]{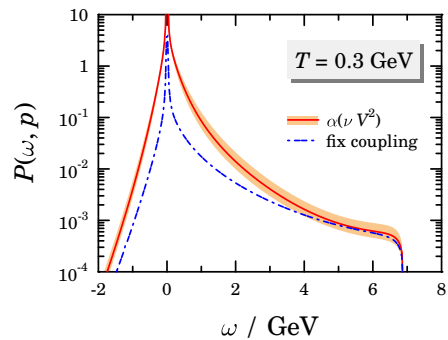}}
	\vskip -5mm
	\caption{Probability density $\sum_{i=g,q} P_i(\omega,p = 8\,$GeV). The band gives the sensitivity under variation $\nu \in [\frac12,2]$ of the scale in the running coupling $\alpha(\nu\, V^2)$. Shown for comparison is the result obtained with the coupling $\alpha_{\rm fix}$ and cut-off $\mu_{\rm fix}$.
 	\label{fig: Pomega}}
\end{figure}

Returning to Eq.~\eq{eq:dEdx}, the thermal 2-body phase space can be expressed as an integral over $\omega$ and the invariant momentum transfer $t$ \cite{Peigne:2007sd}.
This allows us to write
\[
  \frac{dE_i}{dx}
  =
  v^{-1} \int d\omega\, P_i(\omega, p)\, \omega \, ,
\]
with the probability density 
\footnote{Note that only the first moment of $P(\omega)$ is logarithmically enhanced, which justifies the parameterization \eq{eq: mu2}.}
\be
  P_i(\omega, p)
  =
  \int_k \frac{n_i(k)}{2k}\, \bar n_i(k+\omega)
  \int_{t_-}^{t_+}\! \frac{dt}{\sqrt{H}}\, \sum\l|{\cal M}_i\r|^2
\label{eq:P}
\ee
for a test particle with momentum $p$ to change, per unit of time, its energy by $\omega$. 
The condition $H \ge 0$, where
\bea
	H 
	&=& (4\pi)^4 E^2 \l[
	\l(s-(E+k)^2 \r)t^2
	+
	\l( (2Ek-s+m_c^2)^2 \r. \r.
	\nonumber \\
	&& \quad
	\l. - 4k^2p^2+2\omega(k(s+m_c^2)-E (s-m_c^2))  \r)t
	\nonumber \\
	&& \quad
	 \l. - \omega^2 (s-m_c^2)^2 \r] ,
\eea
determines not only the bounds $t_\pm$ in Eq.~\eq{eq:P}, it also constrains the integral over $\bm k$, in particular such that $k+\omega \ge 0$ is always fulfilled. Thus energy gain is exponentially suppressed for $\omega \ksim -T$. For $\omega > 0$, $P \sim \omega^{-2}$ up to near the kinematic threshold, see Fig.~\ref{fig: Pomega}, which also shows the uncertainties arising from the scale setting.

\medskip

In the following we consider an ensemble of test particles. In order to describe its momentum distribution $f(p)$ it is convenient to introduce the rate ${\cal P}(p',p)$ for transitions $p \to p'$. 
The spectrum changes by scattering out and into a given state: for a sufficiently small time interval $f(q, \delta t)	= \l( 1 \!-\! \delta t\, \Gamma(q) \r) f(q) + \delta t \!\int\!dp\, {\cal P}(q,p)\, f(p)$, where $\Gamma(q) = \int\!dp\, {\cal P}(p,q)$ denotes the interaction rate of particles with momentum $q$. Introducing 
\be
	{\cal T}(q,p)
	=
	\l( 1 - \delta t\, \Gamma(q) \r) \delta(q-p) + \delta t\, {\cal P}(q,p)
	\label{eq: Tqp}
\ee
makes explicit that the time evolution of the spectrum,
\be
	f(q, t+\delta t)
	=
	\int\!dp\, {\cal T}(q,p)\, f(p, t) \, ,
	\label{eq: Markov}
\ee
is a first order Markov process (the transition depends only on the state rather than on preceding history).
We discuss the resulting Markov chain on a discrete momentum space, which arises naturally from binning \footnote{Binning regulates the divergence of the interaction rate $\Gamma(q) = \int\!dp\, {\cal P}(p,q)$ arising from $p \simeq q$.} and reduces the convolution \eq{eq: Markov}  to a matrix multiplication,
\be
	f_q(t+\delta t)
	=
	{\cal T}_{q p} \, f_p(t) \, .
\ee
In other words, the evolution in (discrete) time is determined simply by powers of the transition matrix ${\cal T}_{q p}$, which is not only conducive to numerical studies, it also brings instructive insight readily \footnote{It will give a supplementary perspective on conclusions derivable from Boltzmann's equation.}.
Indeed, essential properties of the evolution follow alone from 
\be
	\T\sum_q {\cal T}_{qp} = 1 \quad \mbox{for all \ } p \, ,
\ee
which holds by definition of $\Gamma(q)$.
It implies first that ${\cal T}_{qp}$ maps the hyperplane $\bm H$ characterized by $\| f \| \equiv \sum_p f_p = $ constant onto itself.
Put differently, the `norm' $\| f \|$ is invariant under ${\cal T}_{qp}$, which in the present context describes particle number conservation.
From this we infer $|\lambda_i| \le 1$ for the eigenvalues of ${\cal T}_{qp}$; otherwise the norm would not be conserved in repeated mappings.
Among $\{ \lambda_i \}$ is an eigenvalue $\lambda_1 = 1$ since the rank of the matrix ${\cal T}_{qp} - \bm 1$ is $\dim({\cal T}_{qp})-1$.
These properties of the eigenvalue spectrum are proven more rigorously in the Perron-Frobenius theorem, cf.~\cite{matrix}, which also shows that $\lambda_1$ is a {\sl simple} eigenvalue. Consequently, the stationary state
\be
	f_q^{\rm eq}
	=
	\lim_{n \to \infty} {\cal T}_{qp}^n\, f_p(0) \, ,
\ee
is unique and approached for each initial distribution.

\begin{figure}[ht]
   \centerline{\includegraphics[width=3.2cm]{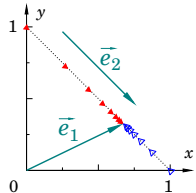}}
	\vskip -5mm
	\caption{Equilibration in 2 dimensions (where $\bm H$ is given by $x+y=1$); the initial states are (1,0) and (0,1), respectively.
 	\label{fig: 2d}}
\end{figure}

The corresponding eigenvectors $\bm e_i$ of ${\cal T}_{qp}$ form a basis with the peculiarity that only $\bm e_1$ has a component perpendicular to the hyperplane, \ie, $\bm e_i \in \bm H$ for $i \not= 1$, see Fig.~\ref{fig: 2d}.
Hence, we can separate in the evolution
$
	f(n\, \delta t)
	=
	\sum_{i=1} \lambda_i^n\, \phi_i \bm{e}_i
	=
	 f^{\rm eq} + \sum_{i=2} \lambda_i^n\, \phi_i \bm{e}_i
$
a part which proceeds entirely in $\bm H$.
Since $|\lambda_{i \not = 1}| < 1$, the approach to equilibrium for large $t$ is exponentially fast,
\be
	\| f(t) - f^{\rm eq} \|
	\; \propto\;
	\exp(-t/\tau^{\rm mix}) \, ,
\ee
where the so-called mixing time
\be
	\tau^{\rm mix}
	=
	\delta t / \ln(\lambda_2^{-1})
	\label{eq: tau_mix}
\ee
is determined by the second largest eigenvalue \footnote{This consideration gives a transparent interpretation of the relaxation time in the homonymous approximation.}.

\begin{figure}[ht]
	\vskip -2mm
	\centerline{\includegraphics[width=6.3cm]{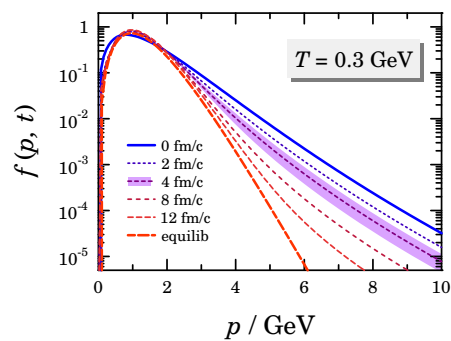}}
	\vskip -5mm
	\caption{Evolution of the initial charm spectrum $I$ at fixed $T$ (for uncertainty band cf.\ Fig.~\ref{fig: Pomega}).
 	\label{fig: f(t)}}
\end{figure}

It is instructive to illustrate these essential features of thermalization  by means of a toy model, namely
\be
 T_{ij} 
	=
	(1-l-g)\, \delta_{i,j} + l\, \delta_{i,j+1} - g\, \delta_{i,j-1} \, ,
\ee
with constants $l$ and $g$ parameterizing the relevant loss and gain rates in ${\cal T}_{qp}$.
First, it is easy to see that the stationary distribution is exponential, $f_i^{\rm eq} \propto \exp(-i/\Theta)$, with the `temperature' $\Theta = 1/\ln(l/g)$ being determined by the logarithmic ratio of loss and gain \footnote{The charm transfer matrix ${\cal T}_{qp}$ will of course lead to equilibration at the temperature of the heat bath, see Fig.~\ref{fig: f(t)}.}.
Furthermore, for large dimension of $T_{ij}$, one can readily derive
\be
	\lambda_2
	=
	1 - ( \sqrt{l}-\sqrt{g} )^2 \, ,
\ee
which shows a non-analytic behavior of the mixing time.

\begin{figure}[ht]
	\vskip -2mm
	\centerline{\includegraphics[width=6.3cm]{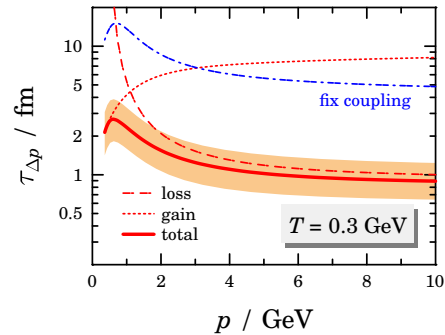}}
	\vskip -5mm
	\caption{Average time $\tau_{_{\Delta p}}$ for a charm quark to change (gain or lose) momentum by $\Delta p > 0.4\,$GeV, compared to the fixed-coupling estimate (uncertainty band as in Fig.~\ref{fig: Pomega}).
 	\label{fig: Gamma}}
\end{figure}

In QCD, for sufficiently  fast charm quarks, the (binned) loss rate dominates over gain (cf.\ Fig.~\ref{fig: Gamma}), thus
\be
	\tau^{\rm mix}
	\,\sim\,
	\Gamma_{\rm loss}^{-1}
	\,\sim\,
	\Gamma_ {\rm total}^{-1} \, .
\ee
This relation between long and short-time aspects of equilibration is worth emphasizing.
More important, Fig.~\ref{fig: Gamma} reveals that conventional calculations would overestimate this crucial time scale by a factor $K \approx 5 \simeq \kappa^{-1}$. 

Realizing that binary collisions are far more effective than previously estimated, we consider in an exploratory study jet quenching in the mid-rapidity region of central collisions within the Bjorken model.
\begin{table}[hb]
\begin{tabular}{c||c|c|c|c|c|c}
set	& $T_0$  & $T_c$ & $\tau_0$ & $t_{\rm life}$ & $R$ & $dN_{\rm ini}/dp_t^2$
\\ \hline
$I$ 	& 0.42$^\star$ 	& 0.18 	& 0.6 			& 7.4 &  5.0$^\star$ & $(p_t+0.5)^2(1+p_t/6.8)^{-21}$
\\
$II$ & 0.30 	& 0.165 	& 1.0 			& 5.0 &  6.6$^\star$ & $(p_t^2+1.8^2)^{-3.5}$
\end{tabular}
\caption{Representative parameter sets for the Bjorken model
(units: GeV or fm; $^\star$adjusted to total entropy $S \approx 10^4$).}
\label{tab: Bjorken params}
\end{table}
To compare with existing results \cite{vanHees:2004gq, Moore:2004tg}, we adopt  parameterizations as summarized in Tab.~\ref{tab: Bjorken params}. For transparency, we will not discuss effects of hadronization (for heavy flavors the partonic suppression can be indicative for the observed $R_{AA}$).

\begin{figure}[ht]
	\vskip -2mm
	\centerline{\includegraphics[width=6.3cm]{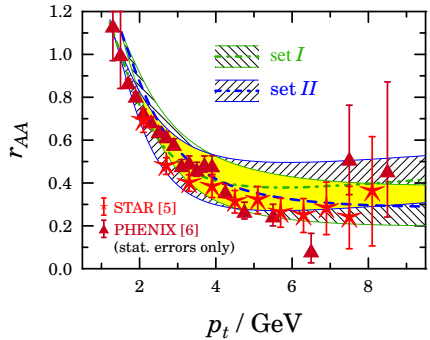}}
	\vskip -5mm
	\caption{Comparison of partonic charm quenching in Bjorken model to the observed $R_{AA}$. For better visibility, the intersection of  uncertainty bands (cf.\ Fig.~\ref{fig: Pomega}) was not hatched.
 	\label{fig: rAA}}
\end{figure}

Consider first the evolution of $f_0(p_t) \propto p_t\, dN_{\rm ini}/dp_t^2$ at constant temperature. 
Fig.~\ref{fig: f(t)} shows the hard part of the spectrum being quenched markedly already after a few fm/c, as could be anticipated from $\tau_{\Delta p} \sim 1$fm/c.
Taking then into account the path length distribution $dN/dl \propto ( 1-(l/2R)^2 )^{1/2}$ in Bjorken's heat bath, $T = T_0(\tau/\tau_0)^{-1/3}$, is straightforward in the Markov formalism. 
Fig.~\ref{fig: rAA} shows the charm quenching ratio $r_{AA} = (dN(t_{\rm life})/dp_t^2)/(dN_{\rm ini}/dp_t^2)$; to compare to $r_{AA} \gsim 0.85$ with fixed coupling, and to Refs.\ \cite{Mustafa:2003vh, vanHees:2004gq, Moore:2004tg, Wicks:2005gt}.
Given that we have calculated the underlying mechanism from first principles, the comparison with the RHIC data \cite{Abelev:2006db, Adare:2006nq} is striking.
Depending on details of the initial spectrum and hadronization,
there is room for radiative quenching (with reasonable parameters), which becomes more important at larger $p_t$.
With regard to the increasing $r_{AA}$ at smaller $p_t$ we underline that the Cronin effect has {\sl not} been taken into account in the calculation.
The small value of $r_{AA}$ at $p_t \gsim 4\,$GeV is quite robust, as verified here with two rather different parameter sets within the Bjorken model. Radial expansion will reduce the life time of the plasma phase, but also modify the path length distribution $dN/dl$. The resulting compensation will be quantified in a forthcoming study with more realistic collision dynamics and including hadronization \cite{GPA}.

\smallskip

To summarize, we have shown in a Markov formalism that, for moderately large momenta, binary collisions are a key mechanism for jet quenching of heavy flavors.
This conclusion requires taking into account -- more carefully than in existing approaches -- essential features of QCD, namely the momentum dependence of the strong coupling and relevant screening effects. Both aspects are interconnected and, as a matter of fact, mandatory for quantitative estimates from thermal field theory, unless temperatures are asymptotically large.

In closing, since jet quenching has significantly influenced the current picture of the strongly coupled quark-gluon plasma (sQGP), it seems worthwhile taking a more general point of view. Commonly used but rather crude estimates (as for the relevant screening range) can lead to interpretations of a `too' strongly coupled plasma.
While for heavy ion phenomenology the relevant coupling is certainly not small, we do see the possibility to understand essential features of the quark-gluon plasma from first principles -- perturbative QCD can work like a charm.

\smallskip

{\bf Acknowledgments:} I thank J.~Aichelin, P.\,B.~Gossiaux,  A.~Smilga and in particular S.~Peign\'e for fruitful discussions.

\end{document}